%% file: article_AIperception_culture.tex
\documentclass[12pt]{article}

\usepackage{arxiv}
\usepackage{setspace}
\usepackage[utf8]{inputenc} % allow utf-8 input
\usepackage[T1]{fontenc}    % use 8-bit T1 fonts
\usepackage[nolist]{acronym}
\usepackage{hyperref}   
\usepackage{graphicx}
\usepackage{csquotes}
\usepackage{microtype}
\usepackage{censor}
\usepackage{rotating}
\usepackage{rotating}
\usepackage{pdflscape}
\usepackage{array}
\usepackage{booktabs}
\usepackage{threeparttable}
\usepackage{threeparttablex}
\usepackage{multirow}
\usepackage{longtable}
\usepackage{xcolor}
\usepackage{doi}
\usepackage[backend=biber,style=authoryear-comp,sorting=ynt]{biblatex}
\usepackage{hyperref}
\hypersetup{colorlinks=true, linkcolor=blue, urlcolor=cyan}

\title{Cultural Dimensions of AI Perception:\\Charting Expectations, Risks, Benefits, Tradeoffs, and Value in Germany and China}

\renewcommand{\shorttitle}{Charting AI Perceptions Across Cultures in Germany and China}
\date{} % This will hide the date

\author{
\href{https://orcid.org/0000-0003-2837-5181}{\includegraphics[scale=0.06]{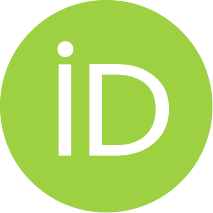}\hspace{1mm}Philipp Brauner} \\
	Communication Science\\
	RWTH Aachen University\\
	52056 Aachen, Germany\\
\And
\href{https://orcid.org/0000-0001-7054-8520}{\includegraphics[scale=0.06]{orcid.pdf}\hspace{1mm}Felix Glawe} \\
	Communication Science\\
	RWTH Aachen University\\
	52056 Aachen, Germany
\And
\href{https://orcid.org/0000-0001-5065-1619}{\includegraphics[scale=0.06]{orcid.pdf}\hspace{1mm}Gian Luca Liehner} \\
	Communication Science\\
	RWTH Aachen University\\
	52056 Aachen, Germany
\And
\href{https://orcid.org/0000-0002-9030-6999}{\includegraphics[scale=0.06]{orcid.pdf}\hspace{1mm}Luisa Vervier} \\
	Communication Science\\
	RWTH Aachen University\\
	52056 Aachen, Germany
\And
\href{https://orcid.org/0000-0002-6105-4729}{\includegraphics[scale=0.06]{orcid.pdf}\hspace{1mm}Martina Ziefle} \\
	Communication Science\\
	RWTH Aachen University\\
	52056 Aachen, Germany
}

\begin{document}

\begin{acronym}
	\acro{ai}[AI]{Artificial Intelligence}
	\acro{ml}[ML]{Machine Learning}
	\acro{dl}[DL]{Deep Learning}
	\acro{ca}[CA]{Conjoint Analysis}
	\acro{tam}[TAM]{Technology Acceptance Model}
	\acro{vam}[VAM]{Value-Based Adoption Model}
	\acro{anova}[ANOVA]{Analysis of Variance}
	\acro{mancova}[MANCOVA]{Multivariate Analysis of Covariance}
\end{acronym}

\maketitle

\begin{abstract}
As artificial intelligence (AI) continues to advance, understanding public perceptions---including biases, risks, and benefits---is essential for guiding research priorities and AI alignment, shaping public discourse, and informing policy.
This exploratory study investigates cultural differences in mental models of AI using 71 imaginaries of AI’s potential futures.
Drawing on cross-cultural convenience samples from Germany (N=52) and China (N=60), we identify significant differences in expectations, evaluations, and risk-benefit tradeoffs.
Participants from Germany generally provided more cautious assessments, whereas participants from China expressed greater optimism regarding AI’s societal benefits.
Chinese participants exhibited relatively balanced risk-benefit tradeoffs ($\beta=-0.463$ for risk and $\beta=+0.484$ for benefit, $r^2=.630$).
In contrast, German participants placed greater emphasis on AI’s benefits and comparatively less on risks ($\beta=-0.337$ for risk and $\beta=+0.715$ for benefit, $r^2=.839$).
Visual cognitive maps illustrate these contrasts, offering new perspectives on how cultural contexts shape AI acceptance.
Our findings highlight key factors influencing public perception and provide insights for aligning AI with societal values and promoting equitable and culturally sensitive integration of AI technologies.
\end{abstract}

\keywords{artificial intelligence, mental models, risk perception, risk-benefit tradeoffs, alignment, technology acceptance, value, cross-cultural differences}

\section{Introduction}\label{section:introduction}

While the origins of \ac{ai} date back several decades \parencite{McCarthy2006,Hopfield1982,Rumelhart1986}, development accelerated due to increased computing power, availability of training data \parencite{Deng2009}, improved algorithms and \ac{dl} \parencite{Lecun2015}, and a surge in funding \parencite{statistaAI2022}.
Many real-world-applications emerged and examples of \ac{ai} applications include its use in education \parencite{Chen2020education}, healthcare \parencite{Amunts2024}, journalism \parencite{Diakopoulos2019}, farming \parencite{Holzinger2024}, as well as manufacturing and production \parencite{Brauner2021}.
All of which promise societal benefits in terms of efficiency, innovation, and convenience but, at the same time, raise fundamental concerns, including risks related to privacy infringement, job displacement \parencite{Acemoglu2017}, and ethical challenges for both individuals and society as a whole \parencite{Awad2018}.

While some view \ac{ai} as a powerful transformational technology that will enhance our lives \parencite{Brynjolfsson2014,Makridakis2017}, others are concerned about its ethical implications and the associated risks \parencite{Cath2018,Bostrom2003}.
Consequently, \ac{ai} currently holds a unique yet ambiguous position in human existence, making it essential to thoroughly investigate how people perceive it, as well as the risks and benefits they associate with it.

The emergence of ChatGPT at the end of 2022 and just recently Deepseek triggered unprecedented public and media interest in \ac{ai}, sparking an explosion of research on how AI is perceived and understood \parencite{Hu2023chatgpt}.
Despite the growing body of research, our understanding of how AI is perceived and its implications for individual lives, societal integration, and governance frameworks remains fragmented and incomplete. A critical gap lies in addressing the so-called \enquote{alignment problem}, which refers to the challenge of aligning AI with individual norms and values \parencite{Gabriel2020, Hristova2024}.
A key limitation is that much of the existing research focuses either narrowly on specific AI applications, usage scenarios and distinct user populations or on very broad philosophical principles.
However, AI perceptions are highly heterogeneous, shaped not only by usage domain and within-country factors like education, personality traits, and socio-economic conditions \parencite{Moravec2024,Castelo2021,Bozkurt2023}, but also by cultural contexts.
As values differ across culture \parencites{Hofstede1984,Jan2022}, cultural differences may play a pivotal role in shaping societal interpretations of \ac{ai}, influencing how individuals and societies perceive, adopt, and desire to regulate AI technologies.
These variations underscore the need for a comparative approach to better understand how diverse cultural backgrounds shape AI perceptions and inform strategies for equitable AI governance.

To address this gap, the present study examines the similarities and differences in AI perception using two samples from Germany and China.
Both countries are major global economies and at the forefront of AI development \parencite{IMF2024, TortoiseMedia2023}.
However, both countries feature unique and diverse cultural values and foci:
Germany's individualistic culture, with a focus on personal privacy and autonomy, may lead to a more critical view of AI \parencite{Hofstede1984}.
In contrast, China's collectivist culture, with a greater emphasis on community and societal harmony \parencite{Hofstede1984}, may contribute to a more favorable view of AI, especially if it is seen as serving the greater good.
Also, China's long-term strategic planning, as seen in initiatives like \enquote{Made in China 2025} may create higher public expectations that AI will be a cornerstone of future success \parencite{madeInChina2025}.
Germans generally exhibit lower acceptance of various technologies compared to Chinese citizens.
This trend is consistent across multiple technologies \parencite{Rieger2021}, including facial recognition \parencite{Kostka2021} and smart homes \parencite{MeyerWaarden2022}.
Also, Germans are more concerned about privacy risks associated with technologies like smart homes, which negatively impacts their intention to use such technologies \parencite{MeyerWaarden2022}.
% In the context of innovation processes in companies, a study suggests that technology perception in Germany is influenced by uncertainty avoidance, while Chinese perception is shaped by power-distance and face-saving concepts \parencite{Tsao2015}.
There is stronger support for government control of technologies in China compared to Germany, reflecting the political contexts of the two countries \parencite{Rieger2021}.

We investigate how participants assess a wide range of potential AI capabilities: how likely they believe these capabilities are to emerge, the risks and benefits they attribute to them, and whether their overall evaluation of these projections is positive or negative. Additionally, 
we analyse the results in terms of similarities and differences between the participants from China and Germany and illustrate the different expectancies and evaluations on visual maps of the queried topics.
Overall, this article provides insights into how various potential applications of AI are perceived and ranked in terms of expectancy, overall evaluation, and risk and benefits perception, as well as how these perceptions are influenced by cultural socialization.

The article’s structure is as follows.
Section \ref{section:relatedwork} reviews presents the related work on public perception of AI and on examines cultural differences in technology perception.
Section \ref{sec:method} describes our methodological approach, the design of our survey, and the two samples of participants from Germany and China.
Section \ref{sec:results} reports present the results of the study, and Section \ref{sec:discussion} analyzes discusses these findings in depth, along with the limitations of the study.
Section \ref{sec:conclusion} summarizes this article and provides suggestions for future research and governance measures.

\section{Related Work}\label{section:relatedwork}

The related work is structured as follows:
First, we introduce the psychometric model used for measuring subjective risk and benefit perceptions within our study which is widely adopted in the field of technology perception in general.
Second, we provide a brief overview on studies on the public perception of AI.
And third, we present studies on culture differences in AI perception, with a focus on works addressing Germany and China.

%% methodological approach
\subsection{Risk Perception and the Psychometric Model}
Individual risk and benefit perceptions influence attitudes, usage intentions, or actual behaviors \parencite{Witte2000,Hoffmann2015,Huang2020}.
Risk can be conceptualised from two perspectives \parencite{Aven2009,Fischhoff1978}:
On the one hand, risk can be modelled as the likelihood of negative events multiplied by the severity of the negative consequences (that is expected values).
However, difficulties arise if probabilities and consequences cannot easily be modelled for complex and multifaceted fields, such as the potential implications of \ac{ai}.
On the other hand, risk can be viewed as events and their perceived consequences and an individual’s perceived threat from these consequences can be measured using subjective rating scales \parencite{Slovic1979}.
This so-called psychometric model offers a framework for understanding how individuals perceive and balance risks and benefits, making it particularly useful for studying perceptions of emerging technologies or technologies whose implications are not easily understood by laypeople \parencite{Fischhoff1978, Slovic1986}.
For example, research explored the role of risk perception in areas such as gene technology \parencite{Connor2010}, technologies for  Carbon Capture and Utilization \parencite{Arning2020}, genetically modified food \parencite{Verdurme2003}, nuclear energy \parencite{Slovic2000}, and climate change \parencite{Pidgeon2011}. 
%Similarly, we assume that laypeople are not fully aware of all the risks and benefits associated with the use of AI. At best, they have a limited and fragmented understanding of the impacts of AI and the resulting changes in society, the job market, or other areas. 

A commonly observed pattern across all topics is the inverse relationship between perceived risks and perceived benefits \parencites{Alhakami1994,Efendi2021}:
When a technology is perceived as risky, its benefits or utilities are often viewed as less significant, whereas technologies perceived as safer tend to be associated with higher perceived benefits, and vice versa.

\subsection{General Public Perception of AI}

Research indicates a dynamic and multifaceted landscape of attitudes and perceptions of \ac{ai} that are shaped based on time, usage context, cultural background, and individual differences.
Given the breadth of tasks and domains affected by \ac{ai}, providing an exhaustive account of public perceptions, especially concerning risk-benefit tradeoffs, is challenging.
Further, the unprecedented adoption of ChatGPT following its public release in 2022 \parencite{Hu2023chatgpt} has further intensified scholarly interest in \ac{ai}.
Consequently, we will give a brief but necessarily incomplete overview of this rapidly evolving field that has yet to be consolidated.

For instance, \textcite{Fast2017} examined three decades of AI coverage in the New York Times, highlighting a marked increase in public interest in AI after 2009.
Their findings revealed a blend of optimism and concern, with AI generally portrayed in a more favorable than unfavorable light.
More recently, however, anxieties regarding loss of control and ethical challenges have become increasingly pronounced, contrasting with continued optimism, particularly regarding AI’s potential in healthcare.

Public surveys suggest that general population often lacks an understanding of AI’s technical achievements and limitations, leading to misconceptions about its functions and potential risks \parencite{IpsosAI2022}.
Research also indicates that public concern regarding the ethical use of AI is increasing, particularly as awareness of biased algorithms and discriminatory outcomes has risen \parencite{ONeil2016}.

A study from the UK investigated common narratives about AI and identified four pessimistic and four optimistic themes \parencite{Cave2019}.
It suggests that AI is often linked to anxiety, as only two of eight identified narratives outweighs the benefits over the concerns (such as the idea that AI could make life easier).
Further, participants reported to feel powerless over AI as developments in the realm of AI are driven by government and corporate interests.
Only half of the respondents could provide a plausible definitions of what AI is, while a quarter thought AI is equivalent to robots.

Two different studies addressed the perception of AI in each country individually.
In Germany, public perception of AI is marked by a mix of admiration for its potential and fear of its risks, particularly cybersecurity threats.
Many view \ac{ai} as a \enquote{black box}, leading to uncertainties and biased control beliefs \parencite{Brauner2023}.
Germans tend to have a cautious approach, with lower trust in AI correlating with a perception of higher risks and lower likelihood of positive impacts.
This cautious stance is reflected in the need for promoting AI literacy to facilitate informed decision-making \parencite{Brauner2023}.
\textcite{Cui2019} investigated the link between media usage and perception of AI in China.
The study measured several factors including media usage frequency, risk perception, benefit perception, policy support, personal relevance, and perceived knowledge.
It found a strong public support for AI and the perception of AI's benefits generally outweighs the perceived risks.
In the study, media usage was only linked to benefit perceptions; which may be due government controlled media that portrays AI in a positive light 
Interestingly, personal relevance reduced media influence, fostering more critical perspectives among individuals highly engaged with AI.

\subsection{Culture and AI and Technology Perception}

\textcite{Grassini2023} shows that cultural backgrounds significantly shape attitudes towards AI. For instance, individuals from the UK and USA exhibit distinct patterns in perceiving AI as either a threat or a benefit, influenced by their cultural contexts, age, and gender.
A recent study examined how generative artificial intelligence (genAI) aligns with values across countries \parencite{globig2024perceived}.
Findings showed that people felt genAI fell short of their desired outcomes. Perceptions of alignment differed by region: Western countries were generally more sceptical and perceived greater misalignment, while Eastern countries tended to be more optimistic, aligned, and trusting. Therefore, cultural values such as individualism and techno-skepticism in Western countries like Germany may drive a more cautious and risk-averse attitude towards AI \parencite{OShaughnessy2022}, whilst collectivist cultures like China may exhibit more trust and optimism towards AI technologies \parencite{Kelley2021}. Barnes et al. (2024) concluded that the cultural identity could influences how AI is integrated into self-conception. Western countries, may view AI as infringing on their autonomy and privacy, while collectivists may see AI as an extension of themselves, aiding in conformity and environmental responsiveness \parencite{Barnes2024}.
In a large international study \textcite{Kelley2021} surveyed the public opinion on AI of over 10000 participants from eight countries (Australia, Canada, USA, South Korea, France, Brazil, India and Nigeria).
The study analysed the anticipated societal impact and the respondents’ attitude towards AI in terms of the four key terms \enquote{exciting}, \enquote{useful}, \enquote{worrying}, and \enquote{futuristic}.
Developed countries—such as USA, Canada, and Australia— predominantly expressed worry about AI, alongside with futuristic expectations. Conversely, in developing countries—such as India, Brazil, and Nigeria—excitement about AI’s possibilities was dominant. Korea stood out for its emphasis on AI’s usefulness and future applications. Overall, there was the widespread expectation that AI will significantly impact society, though its implications remained unclear \parencite{Kelley2021}.
This study highlights that AI perception can relate to the country of residence, being in turn a proxy for cultural differences or economic conditions.

Two countries however absent from this study are Germany and China.
Both countries, measured on their GDP, are two of the five largest economies in the world and in a leading role regarding AI research \parencite{IMF2024, TortoiseMedia2023}.
The direction of AI development in those countries builds upon governmental regulations:
Both countries pursue their own specific AI strategy with China’s New Generation Artificial Intelligence Development Plan and Germany's AI Made in Germany strategy \parencite{roberts2021chinese, halterlein2024imagining}.
China's strategy is oriented towards AI's utility and its economy fostering role, with ethics being used as a tool to steer this development.
Germany’s strategy on the other hand portraits AI as a technology with high-risk potential where ethics underscore the importance of human rights.  Additionally, while the German approach is more shaped by discussions and cooperation across multiple sectors, the Chinese strategy is led by the state top down \parencite{tuzov2024two}.
Those diverging strategies might manifest in the public’s perception of AI as well: In Chinese media outlets, AI is presented as a future oriented and innovation fostering technology with an overall positive stance.
In German media, AI is met with relative caution regarding aspects of privacy, fakes or the general impact on daily lives \parencite{qiu2024decoding}.
A study on acceptance and fear of AI comparing a German and Chinese sample by \textcite{Sindermann2022} confirmed a more positive view and higher acceptance of AI in the Chinese Sample compared to the German sample. Interestingly, this difference was not confirmed with regards to fears attached to AI. They further highlighted that personality traits, as measured in form of the big five, were merely associated to AI perception \parencite{Sindermann2022}.

In February 2023, a survey of 17 countries (including China and Germany) assessed public willingness to trust AI across applications, perceptions of AI’s benefits and risks, and expectations for its development, governance, and oversight \parencites{Gillespie2023}.
 They found that while most people recognize potential benefits, many remain ambivalent or distrustful, with concerns about privacy, fairness, and control, and that trust relies heavily on institutional safeguards and clear knowledge.
 The report also tracks shifts since 2020, noting increased awareness and acceptance in Western countries but persistent skepticism about regulatory effectiveness.

A qualitative study by \textcite{dumbach2021adoption} comparing the adoption of AI in German and Chinese healthcare enterprises does emphasize that German companies are more focussed on challenges related to AI use compared to the chine’s companies, which are more focused on development and future research \parencite{dumbach2021adoption}.
This result highlights one way of how public AI perception can manifest into actual behaviour.
Understanding both national AI strategies and the respective public’s perception in terms of their risk-benefit tradeoff regarding AI can a) foster understanding in collaborative scenarios between German and Chinese actors and b) can help adapt national strategies in preparation to international efforts that might take a different path.

Recently, \parencite{Wang2025} conducted a large-scale survey of public perceptions of AI across 20 countries, incorporating Hofstede’s dimensions of national culture.
The study found that AI perceptions were primarily explained by sociodemographic factors, with men, younger individuals, and those with higher levels of education showing more positive attitudes toward AI.
While similar in scope but much larger in scale than our study, it does not specifically examine the interrelationships between perceived risk, perceived benefit, and attributed value.

\subsection{Research Questions}

Different cultures hold varying beliefs, values, and priorities regarding technology. 
What is considered beneficial or harmful, risky or safe, or socially valuable can differ substantially across societies.
If AI technologies are developed based on assumptions from a single cultural context (such as those from OpenAI or DeepSeek), they may face public resistance or even produce unintended harm when implemented elsewhere. 
Understanding these cultural differences is therefore essential for developing culturally informed AI systems and policies that are globally equitable and locally appropriate.
This study explores these themes through the following research questions:

\begin{enumerate}
	\item How do people from Germany and China differ in their mental models of AI, particularly regarding expectations, perceived risks and benefits, and overall attributed value?
	\item To what extent is the perceived value of AI influenced by perceived benefits versus perceived risks?
	\item Do participants from Germany and China differ in how they trade off AI’s risks and benefits when forming an overall evaluation of its value?
\end{enumerate}

\section{Method}\label{sec:method}

The aim of this study is to explore public perceptions of AI-related risks and benefits, and to examine whether individuals from different cultural backgrounds hold similar or divergent expectations, evaluations, and risk-benefit tradeoffs regarding \ac{ai}.
We selected Germany and China as study sites based on convenient access to participant pools in both countries.
While this choice is not grounded in a theoretically driven cross-cultural comparison, it nonetheless allows for a meaningful exploration of how public perceptions of AI may differ between two countries with distinct technological, political, and cultural contexts.
Given the breadth and complexity of the AI field, we did not focus on a single application but instead examined a diverse set of AI-related topics and asked.
Our research was guided by the following hypotheses:

\begin{enumerate}
\item Perceived risks and benefits across the topics are negatively correlated. This pattern is frequently observed in other domains \parencites{Alhakami1994,Efendi2021} and we assume it carries over to AI perception.
\item Perceived risk is negatively associated with overall evaluations of AI, whereas perceived benefits are positively associated.
\item Participants from Germany will exhibit a more critical stance toward AI technology compared to participants from China.
\end{enumerate}

\begin{figure}
  \includegraphics[width=\textwidth]{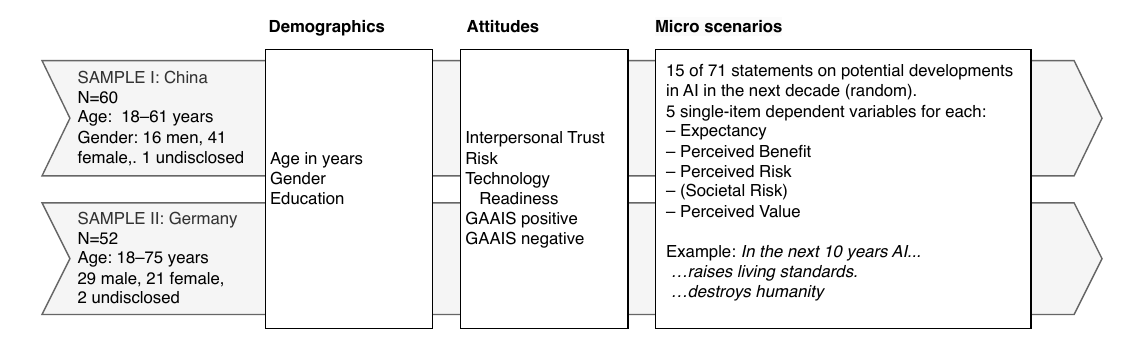}
  \caption{Illustration of the components of the survey and key characteristics of both samples under consideration.}
  \label{fig:studydesign}
\end{figure}

To test our hypotheses, we developed an online questionnaire that captured perceptions of \ac{ai} through a series of imaginary futures (or micro scenarios).
In addition to the constructs relevant to our core hypotheses, we included socio-demographic and individual-level variables based on implicit assumptions and prior research.
The survey was administered in German, English, and Simplified Chinese.
Figure~\ref{fig:studydesign} illustrates our research approach and the following sections describe the design and measured variables in detail.

\subsection{Capturing Perception of AI using Many Scenarios}

The survey asked participants to assess a randomized random subset of 15 out of 71 micro-scenarios \parencite{Brauner2024micro}.
Each micro scenario is a brief statement on potential AI developments, capabilities, or impacts in the next decade that is evaluated on the same four outcome variables.

The list of scenarios was compiled based on previous research and refined through an expert workshop.
We drew on Luhmann’s systems theory (\enquote{Systemtheorie}), which posits that modern society is structured around six essential, functionally distinct, and complementary subsystems: economy, law, science, politics, religion, and education \parencite{luhmann1989ecological,Peterson2004}.
The selected topics spanned a spectrum from more concrete scenarios---such as AI creating jobs or fostering innovation---to more speculative ones, including AI acting on moral principles or perceiving humans as a threat.
Table \ref{tab:tableOfAllItems} shows the complete list of topics.

For evaluating these scenarios, we draw on the psychometric model of risk perception \parencite{Slovic1979,Slovic1985}.
For each topic, participants were asked to rate the likelihood of its occurrence within the next decade (expectancy), assess the personal risk should the projection come true, and evaluate the expected benefits (or perceived utility), each on a single 6-point semantic differential item.
Additionally, we asked the participants to evaluate the societal risk of each statement.
However, we excluded this variable from the analysis due to the very strong correlation with personal risk assessment\footnote{All data is available for interested scholars, see below.}). 
Based on the Value-based Adoption Model by \textcite{Kim2007}, we further included the attributed value (or valence/sentiment), as a target variable to examine how perceived risks and benefits relate to overall evaluations of the technology.
This approach allows us to assess whether the positive or negative value assigned to AI correlates systematically with its perceived risks and perceived benefits.
Using single-item scales is admissible and valid to capture these constructs \parencite{Rammstedt2014,Fuchs2009,Wolfers2025}.

This survey approach provides two distinct perspectives \parencite{Brauner2024micro}:
Firstly, the average evaluations of each participant across all topics evaluated by each participant can be viewed as individual differences (psychometrically, it is a repeated reflexive measurement of the same latent construct).
This allows for the analysis of how participants vary in their evaluations and what influences these evaluations.
Secondly, the average evaluations of each topic across all participants can be seen as a topic factor or topic assessment.
The scores for each topic can then equally be analysed, visually placed on spatial maps, and examined for relationships and patterns between the many topics.

\subsection{Demographics and Individual Level Differences}

Besides the micro scenario evaluations, the survey also queried the participants demographic characteristics and individual-level differences (potentially related personality traits).
For the demographics, we asked for the participants' country of origin, age (in years), and gender (following \textcite{Spiel2019} as \emph{male}, \emph{female}, \emph{diverse}, or \emph{prefer not to say}) as well as their highest educational attainment and current employment.

Since technology can be perceived as a social actor \parencite{Fogg1999,Reeves1996}, we hypothesized that trust would influence the overall evaluation of AI.
Therefore, we measured interpersonal trust using the three-item KUSIV3 scale \parencite{GESISKUSIV3}.
The scale shows strong psychometric properties while being quick to administer at the same time.
We further measured the participants' technology readiness (or technology commitment) using a subset of items from the Technology Commitment Scale \parencite{NeyerZIS244}.
This describes an individual’s tendency to embrace and effectively use new technologies.
We assume that this trait positively influences attitudes towards \ac{ai}.
Additionally, we measured the participants willingness to take or tolerate risks on the single-item Risk Proneness Short Scale (R-1) \parencite{GESISR1}.
We propose that risk proneness is associated with the perceived risk of AI and may negatively impact the overall evaluation of AI.
Lastly, we also administered the General Attitude towards AI scale (GAAIS) with its positive and negative sub scales \parencite{Schepman2022} to study how an overall \ac{ai} assessment relates to the evaluation of many different applications of \ac{ai}.

Except for the scale measuring interpersonal trust (KUSIV3, $\alpha = .540$)\footnote{Low Cronbach’s $\alpha$ values can be acceptable for validated short scales and especially for broader constructs, such asinterpersonal trust, where limited item numbers and conceptual diversity reduce internal consistency without undermining validity.}, all scales demonstrated good to very good internal reliability.
The technology readiness scale had a reliability of $\alpha = .723$, the positive subscale of the General Attitude towards AI (GAAIS-positive) showed very good internal consistency ($\alpha = .800$), and the negative subscale (GAAIS-negative) also exhibited high reliability ($\alpha = .770$).

\subsection{Sample Acquisition}

The survey was available in German, English, and Simplified Chinese.
To ensure consistency across all three languages, native speakers translated the survey.
We used Qualtrics and its multi-language feature for administering the survey.
It started by obtaining informed consent from participants and informing them that participation is voluntary, no personal information will be gathered, and the collected data will be shared as open data.
The study was conducted following protocols approved by our Institutional Review Board (IRB) for studies with the same design with a different sample (ID \censor{2023\_02b\_FB7\_RWTH Aachen}).
All materials, data, and analyses are available as an OSF repository \url{https://osf.io/cpk97/}.

To recruit participants for this survey, we used our personal social networks and asked for forwarding the survey link to others (convenience and snowball sampling).
One Chinese graduate candidate provided significant assistance by disseminating the survey to friends, colleagues, and family members.

\subsection{Data Cleaning and Analysis}

In total 188 people from 28 different countries participated in the study.
Based on our sampling method, the majority came from China and Germany, but we also had participants from India, the United States, and Great Britain.

We filtered the data for non-matching participants (excluding participants from other countries than Germany or China) and incomplete, or low quality responses based based on a) the participant must have completed the full survey b) the participant is not classified as a speeder (less than 1/3 of the median survey duration of 9.6 minutes); the latter is usually considered sufficient for detecting meaningless data \parencite{Leiner2019}).
After filtering, the resulting dataset consists of 112 surveys.
The filter rate of 41.5\% is typical for voluntary and unrewarded surveys, especially, as we had to exclude participants from other countries.

We analyzed the data using R version 4.3.2 with both parametric and non-parametric methods, including Chi-square ($\chi^2$) tests, Bravais-Pearson correlation coefficients ($r$), $t$-tests, Multivariate Analysis of (Co-)Variance (MAN(C)OVA), and Ordinary least squares Multiple Linear Regressions with robust standard errors.
For the omnibus MANOVA/MANCOVA tests, we used Pillai’s trace ($V$) as the test statistic.
We also assessed the assumptions underlying each test and report any violations.
Missing responses were deleted on a test-wise basis.
In line with common practice in the social sciences, we set the Type I error rate at 5\% ($\alpha=.05$) for statistical significance.
All data and associated calculations are publicly accessible through our data repository (see the data availability statement).
 
\subsection{Sample I --- China}

The first sample consists of 60 participants (52.7\% of the whole sample) reporting being from China.
The age of the participants ranged from 18--61 years with a median age of 28 years.
16 participants reported being male, 41 reported being female and 1 person preferred not to disclose their gender identity.
The sample is fairly educated with 50.0\% (29) holding a Bachelor's, 27.6\% (16) holding a Master's, and 10.3\% (6) a PhD degree.

None of the explanatory factors showed a significant association with the demographic variables age or gender.
The largest observed correlation was between age and GAAIS-negative ($r = .343$); however, this correlation was not statistically significant ($p = .083$).
% Age negatively correlation with GAAIS-Neg but not significant
% r=0.343, p=.084>.05

\subsection{Sample II --- Germany}

The second sample consists of 52 participants (47.3\%) reporting being from Germany.
The age ranged from 18--75 years with a median age of 25 years.
29 participants reported being male, 21 reported being female, and 2 preferred not to disclose their gender.
The sample is also quite educated with 57.7\% holding a Bachelor's and 9.6\% (5) a Master's degree and 28.8\% (15) reported having a high school diploma.

Among the explanatory variables, only technology readiness was significantly correlated with gender ($r = .456$, $p = .007$), with women reporting lower technology readiness scores than men.
None of the other explanatory variables were significantly associated with either age or gender.

\section{Results}\label{sec:results}

The results section is structured as follows.
First, we analyse the differences between both samples in terms of the explanatory variables.
Second, we investigate if and how the country of origin influences the overall \ac{ai} expectancy and valuation, as well as the perceived risk and benefits, followed by a brief presentation of the highest/lowest evaluated topics for each sample and dimension.
Next, we analyse how the topic evaluations are interrelated and analyse the risk-benefit tradeoffs for the samples.
Finally, we will explore how individual differences affect the perceptions of AI.

\subsection{Differences between the samples in the explanatory variables}

First, we analyse similarities and differences between both samples in terms of the individual-level differences.
Hereto, we calculated a \ac{mancova} with the sample as independent variable,
age and gender as covariates,
and the individual-level difference variables General Attitudes towards Artificial Intelligence (positive and negative), interpersonal trust, technology readiness, and risk proneness as dependent variables.

Overall, the sample ($V = .280$, $F(5, 95) = 7.405$, $p < .001$) as well as gender ($V = .239$, $F(5, 95) = 5.964$, $p < .001$) have a strong and significant effect on the individual-level differences, with the effect of the sample being stronger than that of gender.
Surprisingly, the sample has a significant effect only on reported technology readiness ($p < .001$), with higher scores for participants from Germany ($M = 4.51$, $SD = 0.60$) compared to participants from China ($M = 4.06$, $SD = 0.50$).
Table \ref{tab:sampledescription} shows the values for each sample.

In contrast, gender has a lower overall effect but significantly influences most individual level differences.
Specifically, genders differ in terms of interpersonal trust, with women reporting higher trust scores ($M = 4.17$, $SD=0.70$) than men ($M = 3.87$, $SD=0.89$), and in technology readiness, where women report lower scores ($M = 4.06$, $SD=0.46$) compared to men ($M = 4.59$, $SD=0.63$).
\begin{table}[htbp]
\centering
\begin{threeparttable}
\label{tab:sampledescription}
\caption{Descriptive overview of the German and Chinese samples.}

\begin{tabular}{llrrc}
\toprule
\textbf{Variable} & \textbf{Measure} & \textbf{Germany (N=52)} & \textbf{China (N=60)} & Sig.\\
\midrule
Age & Mean (SD) & 26.3 (9.4) & 28.6 (8.7) &	$p=.195$\tnote{a}\\
            & Median & 25 & 28 \\
            & Range & 18--75 & 18--61 \\
\midrule
Gender & Male & 29 & 16 	&	$p=.005$\tnote{b}\\
                & Female & 21 & 41 \\
                & Diverse &	0 & 0 \\
                & Prefer not to say & 2  & 1 \\
\midrule
\multirow[t]{6}{*}{Highest Education} & High school & 15 & 1  &	$p<.001$\tnote{c}\\
                   & Apprenticeship & 2  & 4 \\
                   & Bachelor & 30 & 29  \\
                   & Master & 5  & 16 \\
                   & PhD & 0 & 6 \\
                   & Not specified & 0 & 2 \\
\midrule
\multirow[t]{6}{*}{Individual level differences} & GAAIS (positive)  & 4.32 (0.70) & 4.34 (0.50) &	$p=.729$\tnote{b}\\
&	GAAIS (negative)  & 3.48 (0.81) & 3.19 (0.68) &	$p=.056$\tnote{d}\\
&	Interpersonal Trust  & 4.01 (0.83) & 4.03 (0.77) &	$p=.916$\tnote{d}\\
&	Technology Readiness & 4.51 (0.60) & 4.06 (0.50) &	$p<.001$\tnote{d} \\
&	Risk Proneness  & 5.62 (2.01) & 6.15 (1.86) &	$p=.102$\tnote{d} \\
\bottomrule
\end{tabular}

\begin{tablenotes}
\small
\item[a] Welch’s two-sample $t$-test used for unequal variances. No significant age difference between both samples.
\item[b] Pearson’s Chi-squared test for independence indicates a significant gender imbalance, with a higher proportion of women in the China sample.
\item[c] Pearson’s Chi-squared test for independence. The significant education imbalance reflects higher reported education levels in the China sample. Given the convenience sampling, we assume many participants with only a high school degree were currently enrolled in university, mitigating the actual difference between the two samples.
\item[d] MANCOVA with sample and gender as fixed factors and age as a covariate. There was a significant overall effect of sample ($V = .280$, $p < .001$) and gender ($V = .239$, $p < .001$), but not of age ($V = .067$, $p = .245$). The table reports the effects of sample on each outcome variable.
\end{tablenotes}
\end{threeparttable}
\end{table}

\subsection{Cross-Cultural Differences in Overall AI Expectations, Risk and Benefit perceptions, and Attributed Values}

A MANOVA with the sample as independent variable and expectancy, risk, beneift, and overall value as dependent variables shows a significant overall effect ($V=0.415$, $F(1,104)=17.06$, $p<.001$) as well as significant individual differences on three of the four dependent variables.

\begin{figure}
\includegraphics[width=\textwidth]{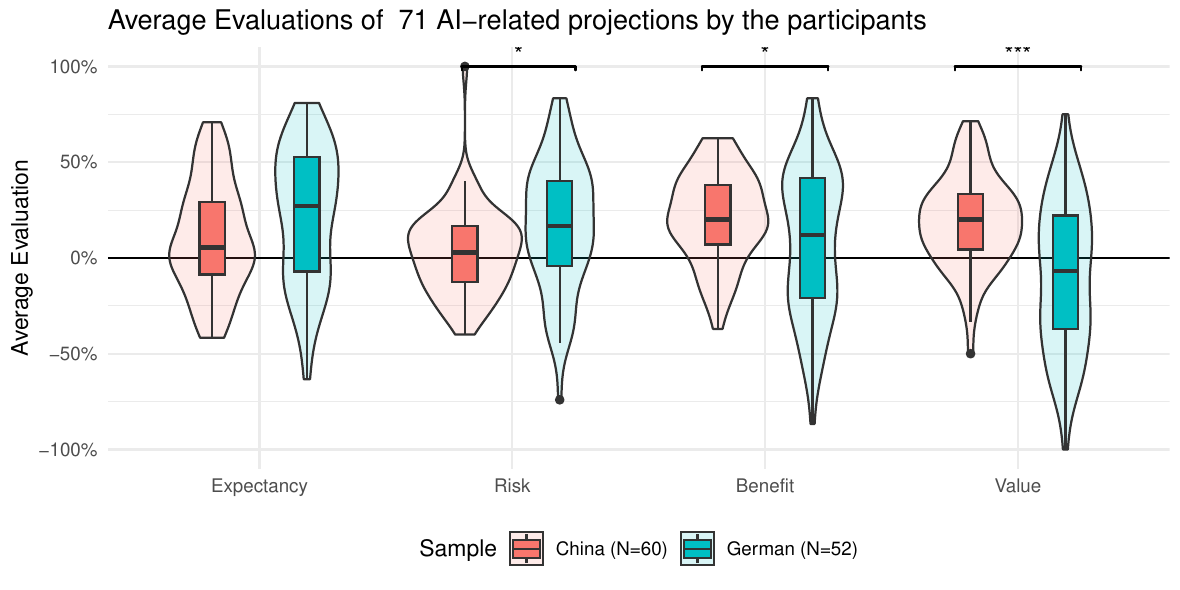}
  \caption{Average evaluation of the 71 micro scenarios on the four assessment dimensions \emph{Expectancy}, \emph{Perceived Risk}, \emph{Perceived benefit}, and overall \emph{Value} with participants from China (N=60) and Germany (N=52). The shaded area below the box plot illustrate the distributions of the individual topic evaluations.}
  \label{fig:violinplot}
\end{figure}

On average, in both samples, the expectation that the statements surveyed will materialize in the next decade is above neutral:
With an average of 11.9\% for the sample from China and an average of 22.2\% for the sample from Germany. 
However, this difference is statistically not significant ($F(1,104)=2.7173$, $p=.102>.05$).
The average risk assessments of the topics of the participants from China is significantly lower (3.5\%) than that of the participants from Germany (18.4\%) ($F(1,104)=2.717$, $p=.021$)).
Further, participants from China evaluate the various projections on the potential impact of \ac{ai} as significantly more beneficial than participants from Germany.
This difference is not only statistically significant but also becomes visually apparent in the analysis of the box and violin plots (\ref{fig:violinplot}):
For the Chinese sample, most of the topics are evaluated from neutral to positive, whereas the distribution of the responses from the German sample are broader and located on both sides of the scale.
Also, participants from China report a higher average perceived benefit (21.5\%) than the participants from Germany (7.1\%) ($F(1,104)=6.502$, $p=0.012$)).
Lastly, the average perceived value (or sentiment) of queried topics is significantly higher for participants from China (20.1\%) than for participants from Germany (-12.4\%) ($F(1,104)=53.628)$, $p<.001$).

Table~\ref{tab:tableDimensionsByCountry} present the average evaluations of both samples and
Figure~\ref{fig:violinplot} shows the average evaluations of all 71 topics per assessment dimension by sample as box plots.
The underlying shadow shows the shape of the data’s distribution.
The stars signify the three significant differences in the four assessment dimensions between participants from China and Germany.

\input{tables/tableDimensionsByCountry.tex}

\subsection{Expectancy}

\begin{figure}
\includegraphics[width=\textwidth]{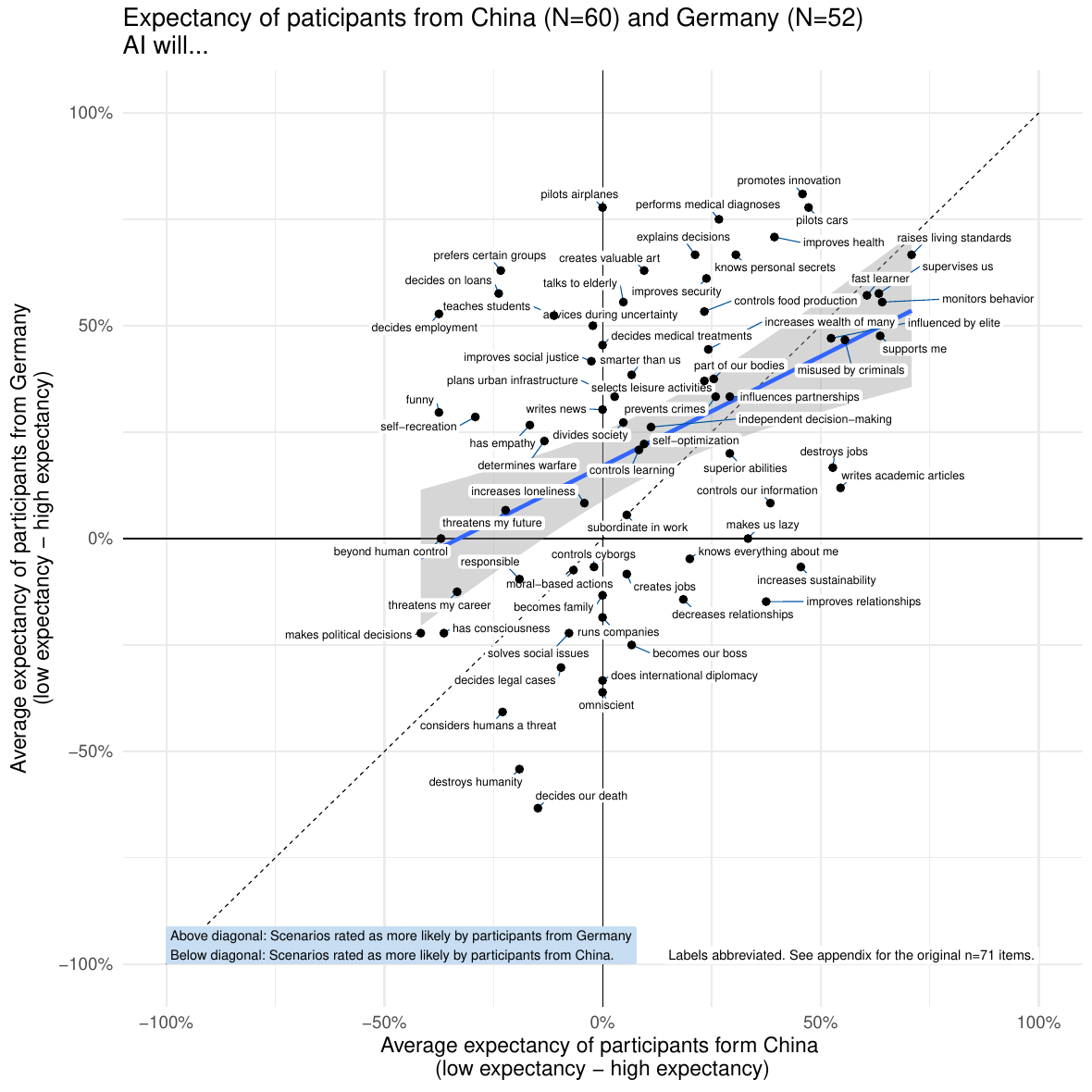}
\caption{Average expectancy ratings for each of the 71 AI-related micro scenarios. The $x$-axis shows the average rating from participants in China (N=60), and the $y$-axis shows the average rating from participants in Germany (N=52). The blue line represents the regression line, with the shaded grey area indicating the 95\% confidence interval. A moderate correlation is observed between the two samples ($r = .419$, $p < .001$). Points that deviate from the diagonal reflect differences in expectancy ratings between the two groups: points above the diagonal indicate scenarios rated as more likely by German participants, while points below the diagonal indicate scenarios rated as more likely by Chinese participants.}
\label{fig:expectancy}
\end{figure}

Although participants from China and Germany do not differ significantly in their overall expectations across all queried topics, there are notable differences in specific AI-related statements.
Figure~\ref{fig:expectancy} illustrates the expectations that the samples from Germany ($y$-axis) and China ($x$-axis) hold regarding AI’s capabilities and implications over the next decade.
Points located on or near the diagonal represent topics where the expectations of participants from both countries are similar.
Points further from the diagonal indicate discrepancies in attributed value:
Points in the upper-left of the diagonal indicate topics that participants from Germany find more likely, while points in the lower-right represent topics seen as more probable by participants from China.

Participants from Germany consider certain AI developments with a significantly higher likelihood than those from China.
For example, they assess that AI will take off, fly, and land airplanes (+77.8\%), independently drive automobiles (+77.8\%), and promote innovation (+81.0\%) as more likely.
In contrast, participants from China believe it is more likely that AI will become a personal assistant that helps solve tasks (63.6\%), supervise public behavior (64.1\%), and raise the standard of living (70.8\%).

German participants assessed it least likely that AI will decide on matters of life and death (-63.3\%), followed by disbelief that AI will destroy humanity (-54.4\%), and that AI will consider humans as a threat (-40.7\%).
Participants from China, on the other hand, assessed it least likely that AI will make political decisions (-41.7\%), followed by disbelief that AI will make decisions in human resources (-37.5\%), and that AI will develop a sense of humor (-37.5\%).

Points with significant differences (as indicated by a strong divergence from the diagonal) include perceptions that AI will favor certain groups (China: -23.3\%, Germany: +63.0\%), make decisions on loan approvals (China: -23.8\%, Germany: +57.6\%), and impact employment decisions (China: -37.5\%, Germany: +52.8\%).
Furthermore, AI is expected to improve relationships (China: +37.5\%, Germany: -14.8\%) and enhance sustainability (China: +45.5\%, Germany: -6.7\%) in notably different ways across the two countries.

\subsection{Value Attributed to AI}

\begin{figure}
\includegraphics[width=\textwidth]{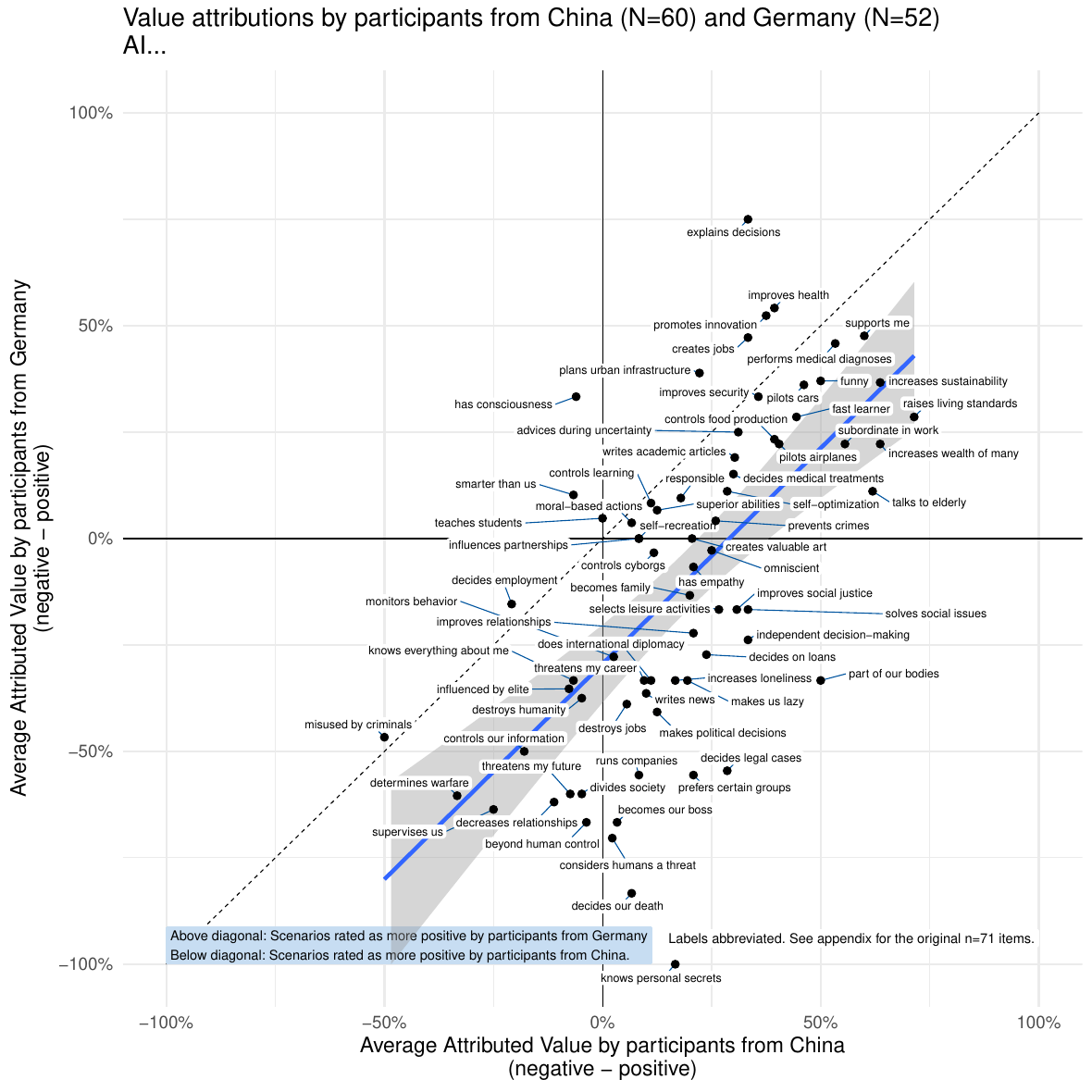}
\caption{Each point represents the average rating of one of the 71 statements by participants from China (N=60, $x$-axis) and Germany (N=52, $y$-axis). The blue line indicates the regression line, with the shaded grey area representing the 95\% confidence interval. While ratings are strongly correlated ($r = .629$, $p < .001$), German participants tend to give lower ratings overall. Points that deviate from the diagonal indicate statements where the two groups differ in their evaluations: points above the diagonal reflect topics rated as more valuable by German participants, while points below the diagonal reflect topics rated as more valuable by Chinese participants.}
  \label{fig:valence}
\end{figure}

Next, we analyze how the two samples differ in the average attributed value towards the \ac{ai}-related projections.
Figure~\ref{fig:valence} presents the average evaluations from the German sample ($y$-axis) and the Chinese sample ($x$-axis).
The interpretation of the points' locations is similar as above.

Again, we do not iterate over each individual statement and instead refer to the detailed data in Table~\ref{tab:tableOfAllItems}.
However, a key observation is that the regression line runs parallel to but significantly below the diagonal, indicating that participants from Germany report a substantially lower overall sentiment toward the queried topics.

Beyond these general trends, certain topics reveal notable differences between the two samples.
Participants from China evaluated the statements that AI may be misused by criminals (-50.0\%), that AI might determine warfare (-33.3\%), and that AI could supervise their private lives (-25.0\%) the least.
For participants from Germany, the statements with the lowest evaluations were that AI knows all secrets (-100.0\%, with all participants rating this item with the lowest possible score), that AI may decide about our lives (-83.3\%), and that AI may consider humans as a threat (-70.4\%).

Participants from China rated the proposition that AI will raise the standard of living the highest (71.4\%), followed by AI increasing the wealth of many people (63.6\%) and making societies more sustainable (63.6\%).
In contrast, participants from Germany evaluated the statement that AI will explain decisions (75.0\%), followed by the expectation that AI will improve our health (54.2\%) and promote innovation (52.4\%) the highest.

Outliers with notable discrepancies between participants from China and Germany include views on whether AI explains its decisions (China: 33.3\%, Germany: 75.0\%), whether AI could develop its own consciousness (China: -6.1\%, Germany: +33.3\%), whether AI will know personal secrets (China: +16.7\%, Germany: -100.0\%), and whether AI will become integrated into human bodies (China: +50.0\%, Germany: -33.3\%).

Figure~\ref{fig:valence} suggests two key points:
First, evaluations show greater alignment compared to expectations, as indicated by the narrower gray area representing the 95\% confidence interval around the regression line.
Second, respondents from China appear to hold a more optimistic view on the statements, derivable by the downward shift in the regression line’s intercept in the figure.
	
\subsection{Perceived Risks}

In the sample from China, participants rated the statement that AI might control food production with a relatively low risk perception (-40.0\%).
Similarly, they attributed low risk to the idea that AI could serve as a conversation partner in elderly care (-33.3\%) or remain subordinate to humans in working life (-33.3\%).
On the other hand, they viewed AI supervising private life (40.0\%) and surpassing human intelligence (40.0\%) as more risky.
The highest risk perception in the Chinese sample was that AI could be misused by criminals, which received a risk score of 100.0\% (\emph{all} responded selected the maximum on the scale).

In the German sample, participants assigned the lowest risk to the notion that AI could become humorous (-74.1\%), followed by the idea that AI might control partner selection (-44.4\%) or make society more sustainable (-40.0\%). 
However, they perceived significant risk in AI supervising private life (72.7\%), acting according to moral concepts (77.8\%), and increasing social justice (83.3\%).

\subsection{Perceived Benefits}

In the Chinese sample, participants attributed the lowest benefit to the idea that AI might supervise private life (-37.0\%), destroy humanity (-28.6\%), or be misused by criminals (-25.0\%).
They viewed AI as offering significant benefit in making society more sustainable (60.6\%), optimizing itself (61.9\%), and deciding who gets an important financial loan (62.5\%).

In the German sample, participants found the idea that AI would know personal secrets to have the least benefit (-86.7\%), followed by AI deciding matters of life and death (-63.0\%) or no longer being controllable by humans (-55.6\%). 
They also perceived low benefit in AI preferring certain groups of people (-55.6\%). 
However, they rated AI’s potential to promote innovation (71.4\%), raise the standard of living (71.4\%), learn faster than humans (81.0\%), and explain its decisions (83.3\%) as highly beneficial.

\subsection{Topic Correlations and the Risk-Benefit Tradeoff}

Next, we analyse how the different evaluation dimensions of the topics are related and if both samples differ in the tradeoff between perceived risks and benefits and what contributes to overall attributed value most.
For this, we calculated a multiple linear regression analysis to examine the relationship between the perceived value of \ac{ai} and the two predictor variables perceived risk and perceived benefit.

First, we calculated a multiple linear regression with perceived risk, benefit and their interaction as well as the sample indicator as predictors and overall attributed value as dependent variable \parencite{Chow1960}\footnote{An alternative approach is to calculate a hierarchical multiple linear regression; which yielded basically the same result. See analysis notebook for reference.}.
The model, including the predictors and their interaction term, explained a significant portion of the variance in value mean ($R^2=.812$, $F(4, 137) = 148$, $p < .001$).
The significance of the sample indicator suggests that the regression models of both samples are significantly different.
Table \ref{tab:regressiontableriskutilitytopics} details the regression model.

Consequently, we calculated two linear regression for each sample with risk, benefit, and the risk-benefit interaction as predictors and value as dependent variables.
For both samples, the models were statistically significant and provided a strong fit for predicting the overall value (sample from Germany: $R^2=.839$, $F(3, 67) = 116.04$, $p<.001$; sample from China: $R^2 = .630$, $F(3, 67) = 38.09$, $p < .001$).

For the sample from Germany, the intercept is negative ($-0.103$) and significant ($p < .001$), suggesting a baseline negative value toward AI statements when risk and benefit are zero.
In contrast, for subjects from China, the intercept is positive ($+0.103$) and also significant ($p<.001$), indicating a baseline positive value.

The coefficient for perceived risk is negative and significant for both the samples from Germany ($\beta=-0.337$) and China ($\beta=-0.463$).
This indicates that as perceived risk increases, overall value attributed towards the AI statements decreases across both countries.
The larger and negative coefficient for China suggests that perceived risk has a stronger negative impact on value for Chinese participants compared to German participants.

The coefficient for perceived benefit is positive and significant ($p<.001$) for both groups: $\beta=+0.715$ for subjects from Germany and $\beta=+0.484$ for subjects from China. This indicates that higher perceived benefit leads to a more favorable overall value towards AI statements.
The effect of perceived benefit is stronger for German participants than for Chinese participants, as shown by the larger coefficient for Germany.

\begin{figure}
\includegraphics[width=\textwidth]{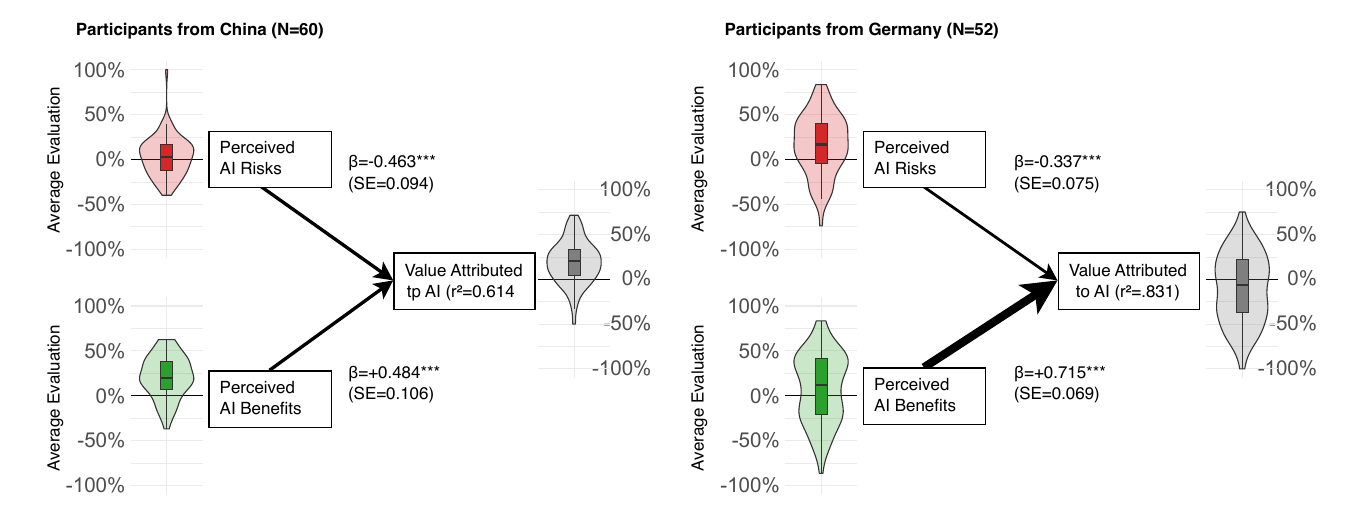}
\caption{Visualization of the two significantly different multiple linear regression models for participants from China (left) and Germany (right), with variable distributions shown as violin and boxplots. In both models, perceived risk and benefit significantly predict attributed AI value. However, German participants placed greater emphasis on benefits, whereas Chinese participants weighed risks and benefits more equally. Line thickness reflects the strength of each predictor’s standardized effect ($\beta$). All predictors are significant at $p < .001$.}
  \label{fig:regression-by-sample}
\end{figure}

\begin{figure}
\includegraphics[width=\textwidth]{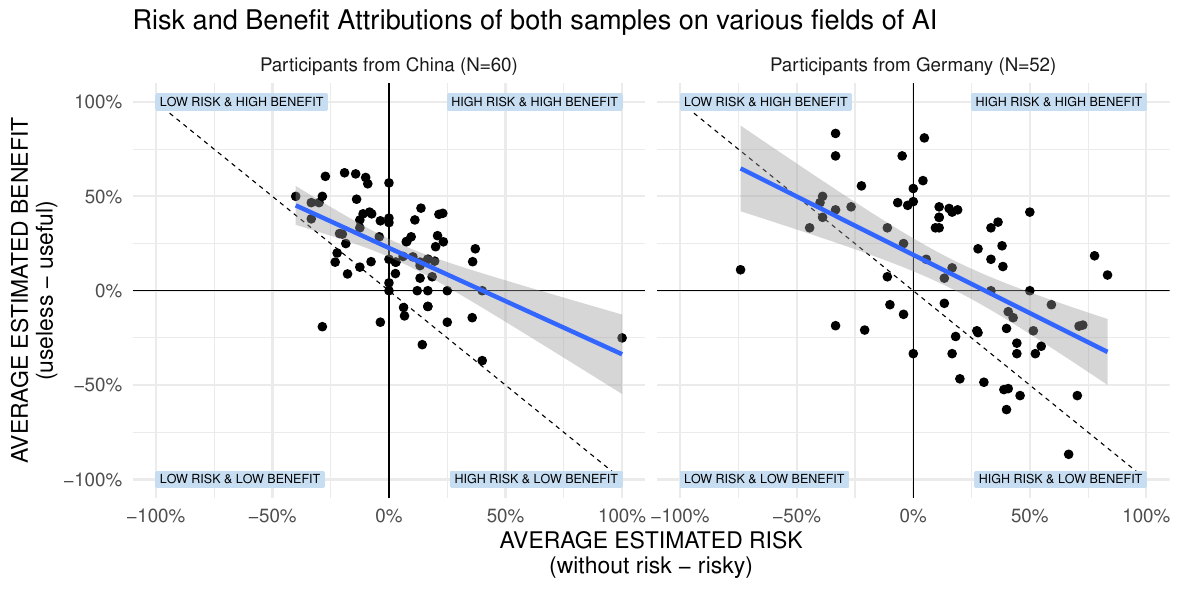}
  \caption{Risk and Benefit attributions by participants from China and Germany on various AI fields. The $x$-axis represents the average estimated risk and the $y$-axis represents the average estimated benefit. The blue lines shows the regression line and the grey area the 95\%-CI of the regression line. The data suggests differing perceptions of AI risks and benefits between the participant from the two countries.}
  \label{fig:riskutility-by-sample}
\end{figure}

The interaction term between risk and benefit is not significant for either country.
This suggests that the combined effect of risk and benefit on attributed value does not significantly deviate from their individual effects.
Table \ref{tab:regressiontableriskutilitytopics} presents the results of all three regression models, while Figure \ref{fig:regression-by-sample} provides a visual comparison.
Figure \ref{fig:riskutility-by-sample} illustrates the different risk-benefit tradeoffs between the participants from China and from Germany by plotting the risk ($x$-axis) and benefit ($y$-axis) evaluations for each topic for both samples.

%% Germany
%In the sample from Germany, the intercept was significant ($I = -0.10$, $SE = 0.029$, $t = -3.61$, $p < .001$), suggesting a baseline negative valence (meaning for neutral risk and benefit perceptions, participants had a slightly negative view) on the \ac{ai} statements.
%Perceived risk was a significant negative predictor of valence mean ($\beta = -0.34$, $SE = 0.07$, $t = -4.80$, $p < .001$), while perceived utility was a strong and significant positive predictor ($B = 0.72$, $SE = 0.07$, $t = 10.58$, $p < .001$).
%The interaction between risk and utility was not statistically significant ($B = 0.09$, $SE = 0.17$, $t = 0.54$, $p = .594$).
%Overall, perceived utility has a stronger positive influence compared to the lower negative influence of perceived risks.
%% China
%For the sample from China,
%the intercept was significant and negative ($B = 0.10$, $SE = 0.03$, $t = 3.69$, $p < .001$) meaning that given a neutral risk and benefit evaluation the valuation of \ac{ai} is slightly negative.
%Risk was a significant negative predictor of \ac{ai} valence ($B = -0.46$, $SE = 0.10$, $t = -4.75$, $p < .001$), while utility was a significant positive predictor ($B = 0.48$, $SE = 0.09$, $t = 5.36$, $p < .001$).
%The interaction term between risk and utility was not statistically significant ($B = -0.02$, $SE = 0.28$, $t = -0.06$, $p = .952$).

\begin{table}[h!]
\centering
\begin{tabular}{rccc}
\toprule
 	  & \multicolumn{3}{c}{Dependent variable: Overall Attributed AI Value}		\\ \cmidrule(l{3pt}r{3pt}){2-4}
Independent Variables & Sample Germany (N=52)	& Sample China (N=60)	& Combined Sample (N=112)\\ \midrule
(Intercept)				& -0.103***		& +0.103***		&	0.070**		\\
(SE)					&	(0.028)		&	(0.027)		&	(0.021)		\\[0.2cm]
Perceived Risk $\beta$	& -0.337*** 	& -0.463*** 		& -0.381***		\\
(SE)					& (0.075)		& (0.094)		& (0.058)			\\[0.2cm]
Perceived Benefit $\beta$& +0.715***	& +0.484***		& +0.646***	 	\\ 
(SE)					& (0.069)		& (0.106)		& (0.054)		\\[0.2cm]
Risk $\times$ Benefit $\beta$ & 0.092	& -0.017 		& 0.141			\\
(SE)					&	(0.167)		& (0.241)		& (0.027)		\\[0.2cm]
Sample $\beta$      & \multirow{2}{*}{---} & \multirow{2}{*}{---} &  -.157***\\
SE               &                    &                    & (0.131)	 \\[0.4cm]
N (topics)				&	71			& 71		&	71	\\
$F$						&	116.0			& 38.1		&	148.0	\\
$p$						&	$<.001$		& $<.001$	&	$<.001$	\\
$R^2$ 					&	.831		& .614		&	.806	\\
\bottomrule
\end{tabular}
\caption{Regression tables for three multiple linear models predicting perceived value of AI statements (interpreted as individual differences) using risk, benefit, and their interaction as predictors. Models are shown for the German sample, the Chinese sample, and the combined sample. The significant Sample variable indicates that the first two models differ significantly. Robust standard errors further suggest significantly different $\beta$-coefficients (rel. influences), reflecting distinct risk-benefit trade-offs between the two samples.}\label{tab:regressiontableriskutilitytopics}
\end{table}

\subsection{The Role of Individual Level Differences on the Perception of AI}

Lastly, we analyse if individual differences influence the assessment of \ac{ai}.
With regards to the associations between the four assessment variables interpreted as an individual difference, both samples show a similar pattern.
First, as for the evaluations of the topics, there is an inverse relationship between perceived risk and benefit in both the sample from China ($r=-.496$, $p<.001$) and Germany ($r=-.342$, $p=.052$).
Second, in both samples risk is associated negatively (China: $r=-.540$, $p<.001$; Germany: $r=-.457$, $p=.003$) and benefit positively (China: $r=.735$, $p<.001$; Germany: $r=.734$, $p<.001$) with the overall value attributed to the topics.
Table \ref{tab:correlationuseroverall} lists the correlation between the demographic variables, the individual-level differences, and the average \ac{ai} attributes.

It is worth noting that many correlations between the individual-level differences and the attributions, though intuitively meaningful and potentially non-negligible, were not statistical significant due to the small sample size; especially in the Chinese sample. Here, for example, correlation between GAAIS-negative and perceived AI Risks was comparatively large but a non-significant ($r>.4$).
We refrain from an deeper analysis of the potential influence of the individual-level differences on the perception of AI and the risk-benefit tradeoff.
Future research with larger sample sizes are needed to address these questions.

\begin{table}[ht]
\centering
\caption{ Correlation of demographic factors and individual-level differences with the AI  attributions interpreted as individual difference. *** signify correlations at $p<.001$, * at $p<.05$, and + at $p<.1$. Other values are $n.s.$.}\label{tab:correlationuseroverall}
\begin{tabular}{lcccccccc}
\toprule
\multicolumn{1}{c}{ } & \multicolumn{4}{c}{Sample China (N=60)} & \multicolumn{4}{c}{Sample German (N=52)} \\
 \cmidrule(l{3pt}r{3pt}){2-5}
 \cmidrule(l{3pt}r{3pt}){6-9}
Variable 				&  1. 	&  2. 		& 3. 		& 4.  		&  1. 	&  2. 	& 3. 		& 4.			\\ 
\midrule
1. Perceived Expectancy & ---	& .653***	& -.116		& -.212 		&	---	& .256	& -.018	&	.017	\\	
2. Perceived Risk		&		&	---	 	& -.496***	& -.541***	&		& ---	& -.342+	&	-.457*	\\
3. Perceived Benefit 	&		&		 	& 	---		& .735***	&		&		& ---		&	.734***	\\
4. Attributed Value 	&		&		 	&			&	---		&		&		& 			&			\\
\midrule
5. Age					& -.013	& -.136		& -.061		& .173		& .160	& -.197 & -.133		& -.002	\\
5. Gender (0=female, 1=male)	& .134	& .203		& -.119		& -.220		& .097	& -.197	& -.006		& .067	\\
6.	Technology Readiness &
.228	&	-.010	&	.157	&	.140	&
-.034	&	-.032	&	.280	&	.177	\\
7.	GAAIS (negative) &	-.001 &	.178	&	-.085 &	-.252	&
	.191	&	.402+	&	-.222	&	-.422* \\
8.	GAAIS (positive) &	.083	&	-.020	&	.053 &	.133	&
.100	&	-.120 &	.454*	& .489*\\
\bottomrule
\end{tabular}
\end{table}

\section{Discussion}\label{sec:discussion}

In this study, we investigated how cultural differences shape the evaluation of \ac{ai}, by studying the perception of \ac{ai} of  participants from Germany and China as an example.
We asked our participants to evaluate 71 statements on the potential capabilities and impacts of AI in terms of the likelihood of occurrence over the next decade, as well as their perceived risks and benefits, as well as the overall value (or sentiment) should these scenarios become a reality.

Regarding the envisioned future of \ac{ai}, the study did not reveal significant differences in the expectations that the presented \ac{ai} scenarios would come true within the next decade.
Although some of the statements were judged differently by participants from both countries, there is no overall difference across all topics.
This is surprising, as other studies suggest that the countries and cultures differ in their openness towards technology \parencite{Espig2022}, which may also carry over to differences in the expectations regarding future technologies. 
% TODO openness != expectancy 
However, this may be due to a methodological weak spot of the study:
Humans are poor at predicting the future \parencite{Recchia2021}.
Hence, asking about the expectations regarding AI might not have been suitable.
% TODO people bad at predicting. but right interpretation?
Second, the selection of scenarios included not only technology-related statements but also social implications of \ac{ai}.
As a result, any potential effect may be diluted due to the diversity of the scenarios.
Consequently, we recommend addressing potential differences in expectations in future, more in-depth studies.

Overall, the results suggest that both perceived risk and perceived benefits significantly influence participants’ value toward AI, with perceived benefits having a stronger positive influence compared to the negative influence of perceived risk.
Both, the inverse relationship between perceived risks and perceived benefits, and that the benefits are a stronger predictor,  is frequently observed in studies on risk perception \parencites{Alhakami1994,Efendi2021}.
This study corroborates these findings and complements that this extends to the AI context and across cultures.

However, the perception of AI significantly diverges between the German and the Chinese sample regarding perceived risk, benefits, and overall value.
The Chinese sample assessed the topics overall more positively than the German sample.
This finding is in line with the previous work of \textcite{Sindermann2022}.
Additionally, as the difference in regression models for both samples shows, perceived benefits had a weaker influence on overall value compared to the German sample coupled with a lower explanation of explained variance.
This leads to the conclusion that the more positive perception of AI in China might be explained by additional factors other than risk and benefit.
However, it is not clear of what nature this factor is:
Demographic factors and individual-level differences, such as age, gender, technology readiness or the general attitude towards AI did not show significant relationships to the AI attributions.
This, again, reflects the outcome of Sindermann's et al. work in which they found only weak associations between personality traits and AI perception.
Multiple previous studies have shown that trust in the own government serves as an important predictor of AI perception: The higher the trust in the government, the more positive the perception of AI \parencite{Chen2020, schmager2024exploring, wang2024citizens}.
Trust in the Chinese government therefore might explain more of the variance of attributed value than perceived risk and benefit alone and should be examined in future studies. 
While we used country of origin as a simplified proxy for culture---acknowledging that culture is a far more complex construct---this contrast, though exploratory, provides valuable initial insights into potential cultural dimensions of AI perception and underscores the need for more systematic, theory-driven cross-cultural research in this area.

% Methodological contribution
In addition to enhancing our understanding of public perceptions of \ac{ai}, this article offers methodological contributions by providing a new perspective on the public perception (of \ac{ai}).
Recent studies indicate that survey results may reflect not only participants’ judgments but also, to a significant extent, the influence of linguistic and lexical properties of surveys, such as word co-occurrences and scale similarities \parencite{Gefen2017}. 
The micro-scenario approach addresses this concern by employing reflexive measurements across a diverse range of topics, rather than relying on psychometric scales of similarly phrased items \parencite{Brauner2024micro}.
The approach yielded plausible findings, suggesting it offers a novel perspective for triangulating cognitive phenomena in technology perception.

\subsection{Limitations}

This study is not without limitations.
First, we relied on convenience and snowball sampling, yielding a small and highly educated academic sample for both countries.
While the modest sample size may limit generalizability, it was gathered through personal networks rather than paid participant pools, fostering more thoughtful and intrinsically motivated responses.
Unlike typical large-scale surveys where participants are incentivized by minimal compensation, our respondents were likely driven by genuine interest or social connection. We believe this approach yielded higher-quality data despite the smaller N.
Although we only sampled a snapshot of the respective populations, we assume that the differences in AI perception will equally emerge in more diverse samples, as both samples are comparable in terms of age (though gender biased), but already exhibit substantial differences in the evaluation of AI's risk, benefits, and overall evaluation.
A related concern is the under-sampling of the individual topics per sub-group owing to the small sample size.
Therefore, the individual evaluations of statements and their placement on the maps should be considered with caution, yet key claims made in the article--- i.e., different overall evaluations in terms of risk, benefit, and value and differences in the risk-benefit tradeoffs---build on aggregations across many statements and across many participants and thus sufficiently many measurements\footnote{Over 6414 evaluations of the statements have been captured and thus over 1200 evaluations per dimension.} (cf. \textcite{Bernoulli1713}).

Second, while culture encompasses a wide range of influences, including values, beliefs, social norms, language, and individual experiences, we solely used the country of origin as a pragmatic proxy for culture \parencite{Schwartz2006}.
Countries typically have dominant cultural frameworks that shape the structures individuals grow up within.
Hence, we conclude that our work contributes to the understanding of cultural differences in technology and \ac{ai} perception, but suggest extending this evidence with a larger, more diverse, and culturally broader sample.
Further, we suggest to include more nuanced cultural aspects in the survey beyond the home country (e.g., Hofstede's (\citeyear{Hofstede1984}) cultural dimensions or Schwartz's (\citeyear{Schwartz2006}) theory of basic values).
This would allow to separate and mitigate the potential influence of contextual factors (such as AI exposure in the different countries) and to better model how culture and its constituting factors affect technology perception and the risk-benefit tradeoffs involved.
This is particularly important in light of recent findings by \textcite{Wang2025}, which suggest that individual-level factors, rather than broader cultural frameworks, primarily shape general perceptions of AI. Whether this also applies to perceived risks and benefits and their trade-offs remains an open question.

Third, the study examined a wide range of AI-related statements across a few dependent variables using micro scenarios \parencite{Brauner2024micro}.
This method captures heuristic evaluations rather than detailed assessments of benefits, barriers, and implications.
As the results show strong and systematic patterns, this suggests reliable measurements and insights into the broader perception of \ac{ai} and the influence of culture.
However, this does not reveal the specific motives behind the evaluations of individual topics.
Given \ac{ai}'s impact on individual lives and society, each topic warrants in-depth qualitative and quantitative studies to enhance our understanding.
This can help practitioners, researchers, and policymakers better align \ac{ai} with human needs.

\section{Conclusion and Outlook}\label{sec:conclusion}

While significant research has addressed AI perception and its societal implications, a critical gap remains in tackling the \enquote{alignment problem}, which seeks to ensure AI systems align with human norms and values \parencite{Gabriel2020, Hristova2024}. 
Given that norms and values vary across cultures and countries, it is imperative to integrate cross-cultural perspectives into AI alignment efforts.

Despite many individual findings on specific topics, our work suggests at a higher level that Germans exhibit a more cautious and risk-aware attitude towards AI, while the Chinese public demonstrates greater optimism, likely influenced by cultural and governmental factors.
Understanding these differences---both for individual use cases and implications as well as the more general perspective---is crucial to better account for the varied norms, values, and ethical frameworks that shape human-AI interactions globally.
Such insights could enhance the adaptability, relevance, inclusivity, and fairness of AI systems across different societal contexts, fostering more equitable and sustainable AI integration worldwide.

Effective alignment and public engagement strategies must consider these cultural differences to address specific concerns and promote responsible AI development tailored to each context \parencite{OShaughnessy2022, Kelley2021, globig2024perceived}.
Additionally, these cross-cultural findings can foster a better mutual understanding and strengthen international collaboration, enabling more effective and culturally sensitive approaches to AI integration and governance.

\clearpage
\section*{Appendix}

\input{tables/itemtable}

\section*{Acknowledgments}

This approach builds on the foundation laid by a prior study from Ralf Philipsen and the first author \parencite{Brauner2023}.
The methodology employed in this study has evolved over multiple studies and extensive discussions with colleagues.
To name just a few, we express our gratitude to Julia Offermann and Julian Hildebrandt for their inspiring viewpoints on the methodological approaches and interpretations.
We thank Tim Schmeckel for research support.
Funded by the Deutsche Forschungsgemeinschaft (DFG, German Research Foundation) under Germany’s Excellence Strategy -- EXC-2023 Internet of Production -- 390621612.

\section*{Declarations}

\subsection*{Funding}
Funded by the Deutsche Forschungsgemeinschaft (DFG, German Research Foundation) under Germany’s Excellence Strategy -- EXC-2023 Internet of Production -- 390621612.

\subsection*{Conflicts of interest}
The authors have no conflicts of interest to declare.

\subsection*{Ethics approval}
The study was conducted following protocols approved by our Institutional Review Board (IRB) for studies with the same design with a different sample (ID \censor{2023\_02b\_FB7\_RWTH Aachen}).

\subsection*{Consent to participate}
Participants were of legal age and provided informed consent for voluntary, uncompensated participation in the study.

\subsection*{Consent for publication}
All authors provided their consent for publication.

\subsection*{Availability of code, data and materials}
All materials, data, and the R based Quarto Notebook of the analysis is available on the Open Science Foundation repository:
\url{https://osf.io/cpk97/}

%\subsection*{authors' contributions}
\printbibliography

\end{document}

%% file: tables/tableDimensionsByCountry.tex
\begin{table}
\centering
\caption{\label{tab:tab:tableDimensionsByCountry}Mean assessments across the four evaluation dimensions were compared between Chinese and German participants. A MANOVA revealed a significant overall effect, with significant group differences for perceived risk, perceived benefit, and attributed value; expectancy did not differ significantly.\label{tab:tableDimensionsByCountry}}
\centering
\begin{threeparttable}
\begin{tabular}[t]{lccccccccc}
\toprule
\multicolumn{1}{c}{ } & \multicolumn{4}{c}{China} & \multicolumn{4}{c}{German} & \multicolumn{1}{c}{Sig.} \\
\cmidrule(l{3pt}r{3pt}){2-5} \cmidrule(l{3pt}r{3pt}){6-9} \cmidrule(l{3pt}r{3pt}){10-10}
Dimension & M & SD & CI-95 & N & M & SD & CI-95 & N & p\\
\midrule
Expectancy & 11.9\% & 62.3\% & {}[-4.1\%,28.0\%] & 58 & 22.2\% & 63.9\% & {}[4.9\%,39.6\%] & 52 & 0.102\\
Perceived Risks & 3.5\% & 55.0\% & {}[-10.6\%,17.7\%] & 58 & 18.4\% & 61.0\% & {}[1.8\%,35.0\%] & 52 & 0.021\\
Perceived Benefits & 21.5\% & 54.1\% & {}[7.6\%,35.4\%] & 58 & 7.1\% & 62.6\% & {}[-9.9\%,24.1\%] & 52 & 0.012\\
Attributed Value & 19.7\% & 51.6\% & {}[6.4\%,33.0\%] & 58 & -11.9\% & 59.1\% & {}[-27.9\%,4.2\%] & 52 & <0.001\\
\bottomrule
\end{tabular}
\begin{tablenotes}
\item \textit{Note: } 
\item Measured on 6-point semantic differential scales, rescaled to a range of -100\% to +100\%. Negative values indicate low evaluations on the respective dimension (e.g., low perceived value, risk, benefit, or expectancy), while positive values indicate high evaluations.
\end{tablenotes}
\end{threeparttable}
\end{table}

%% file: tables/itemtable.tex
\begin{landscape}
\begin{ThreePartTable}
\begin{TableNotes}
\item \textit{Note: } 
\item Measured on 6 point Likert scales and rescaled to -100\% to +100\%. Negative values indicate a negative evaluiation of the respective dimension (i.g., low attributed value, low perceived risk, low perceived benefit, or low expectancy) and positive values indicate a high evaluation. Permission to translate, use, and adapt the items is, of course, granted, provided proper citation.
\end{TableNotes}
\begin{longtable}[t]{>{\raggedright\arraybackslash}p{5cm}rrrrrrrr}
\caption{\label{tab:tableOfAllItems}Question items from the survey and average responses to each item from Chinese (N=60) and German (N=52) participants ordered by the difference in attributed value.}\\
\toprule
\multicolumn{1}{c}{} & \multicolumn{2}{c}{Expectation} & \multicolumn{2}{c}{Perceived Risk} & \multicolumn{2}{c}{Perceived Benefit} & \multicolumn{2}{c}{Attributed Value} \\
\cmidrule(l{3pt}r{3pt}){2-3} \cmidrule(l{3pt}r{3pt}){4-5} \cmidrule(l{3pt}r{3pt}){6-7} \cmidrule(l{3pt}r{3pt}){8-9}
In 10 years, AI... & M China & M German & M China & M German & M China & M German & M China & M German\\
\midrule
\endfirsthead
\caption[]{Question items from the survey and average responses to each item from Chinese (N=60) and German (N=52) participants ordered by the difference in attributed value. \textit{(continued)}}\\
\toprule
\multicolumn{1}{c}{} & \multicolumn{2}{c}{Expectation} & \multicolumn{2}{c}{Perceived Risk} & \multicolumn{2}{c}{Perceived Benefit} & \multicolumn{2}{c}{Attributed Value} \\
\cmidrule(l{3pt}r{3pt}){2-3} \cmidrule(l{3pt}r{3pt}){4-5} \cmidrule(l{3pt}r{3pt}){6-7} \cmidrule(l{3pt}r{3pt}){8-9}
In 10 years, AI... & M China & M German & M China & M German & M China & M German & M China & M German\\
\midrule
\endhead

\endfoot
\bottomrule
\endlastfoot
decides on hiring, promotions, and terminations. & -37.5\% & 52.8\% & 16.7\% & 38.5\% & -8.3\% & 12.8\% & -20.8\% & -15.4\%\\
prefers certain groups of people. & -23.3\% & 63.0\% & -3.7\% & 45.8\% & -16.7\% & -55.6\% & 20.8\% & -55.6\%\\
decides who gets an important financial loan. & -23.8\% & 57.6\% & -19.0\% & 27.3\% & 62.5\% & -21.2\% & 23.8\% & -27.3\%\\
autonomously takes off, flies, and lands airplanes. & 0.0\% & 77.8\% & 21.4\% & -11.1\% & 40.5\% & 33.3\% & 40.5\% & 22.2\%\\
is humorous. & -37.5\% & 29.6\% & -28.6\% & -74.1\% & 50.0\% & 11.1\% & 50.0\% & 37.0\%\\
teaches students. & -11.1\% & 52.4\% & -7.4\% & 9.5\% & 40.7\% & 33.3\% & 0.0\% & 4.8\%\\
can recreate itself. & -29.2\% & 28.6\% & 25.0\% & 38.1\% & -16.7\% & 23.8\% & 8.3\% & 0.0\%\\
creates valuable works of art that are traded for money. & 9.5\% & 63.0\% & 0.0\% & -33.3\% & 16.7\% & -18.5\% & 20.5\% & 0.0\%\\
advises me in uncertain times. & -2.2\% & 50.0\% & -17.8\% & 0.0\% & 8.9\% & 54.2\% & 31.1\% & 25.0\%\\
serves as a conversation partner in elderly care. & 4.8\% & 55.6\% & -33.3\% & -22.2\% & 38.1\% & 55.6\% & 61.9\% & 11.1\%\\
carries out medical diagnoses. & 26.7\% & 75.0\% & -10.0\% & 16.7\% & 60.0\% & 41.7\% & 53.3\% & 45.8\%\\
explains its decisions. & 21.2\% & 66.7\% & 6.1\% & -33.3\% & 18.2\% & 83.3\% & 33.3\% & 75.0\%\\
decides on medical treatments. & 0.0\% & 45.5\% & 20.0\% & 36.4\% & 23.3\% & 36.4\% & 30.0\% & 15.2\%\\
increases social justice. & -2.6\% & 41.7\% & 10.3\% & 83.3\% & 17.9\% & 8.3\% & 30.8\% & -16.7\%\\
has the ability to recognize, understand and empathize with emotions. & -16.7\% & 26.7\% & 0.0\% & 13.3\% & 4.2\% & -6.7\% & 20.8\% & -6.7\%\\
improves the security of people. & 23.8\% & 61.1\% & 7.7\% & 11.1\% & 26.2\% & 38.9\% & 35.7\% & 33.3\%\\
can no longer be controlled by humans. & -37.0\% & 0.0\% & 6.7\% & 70.4\% & -13.3\% & -55.6\% & -3.7\% & -66.7\%\\
determines warfare. & -13.3\% & 22.9\% & 37.0\% & 71.1\% & 22.2\% & -18.8\% & -33.3\% & -60.4\%\\
knows my secrets. & 30.6\% & 66.7\% & 2.8\% & 66.7\% & 9.1\% & -86.7\% & 16.7\% & -100.0\%\\
promotes innovation. & 45.8\% & 81.0\% & -12.5\% & -4.8\% & 33.3\% & 71.4\% & 37.5\% & 52.4\%\\
is more intelligent than humans. & 6.7\% & 38.5\% & 40.0\% & 15.4\% & 0.0\% & 43.6\% & -6.7\% & 10.3\%\\
improves our health. & 39.4\% & 70.8\% & -21.2\% & 4.2\% & 30.3\% & 58.3\% & 39.4\% & 54.2\%\\
determines the construction and infrastructure of our cities. & 2.8\% & 33.3\% & -8.3\% & -38.9\% & 41.7\% & 38.9\% & 22.2\% & 38.9\%\\
independently drives automobiles. & 47.2\% & 77.8\% & 0.0\% & 0.0\% & 38.5\% & 47.2\% & 46.2\% & 36.1\%\\
independently writes news. & 0.0\% & 30.3\% & 12.1\% & 18.2\% & 0.0\% & -24.2\% & 10.0\% & -36.4\%\\
controls food production. & 23.3\% & 53.3\% & -40.0\% & -6.7\% & 50.0\% & 46.7\% & 39.4\% & 23.3\%\\
threatens my private future. & -22.2\% & 6.7\% & -18.5\% & 20.0\% & 25.0\% & -46.7\% & -7.4\% & -60.0\%\\
divides society. & 4.8\% & 27.3\% & -28.6\% & 30.3\% & -19.0\% & -48.5\% & -4.8\% & -60.0\%\\
threatens my professional future. & -33.3\% & -12.5\% & 9.5\% & -4.2\% & 28.6\% & -12.5\% & 9.5\% & -33.3\%\\
increases the wealth of many people. & 24.2\% & 44.4\% & -9.1\% & 11.1\% & 56.7\% & 38.9\% & 63.6\% & 22.2\%\\
makes political decisions. & -41.7\% & -22.2\% & 20.8\% & 59.3\% & 29.2\% & -7.4\% & 12.5\% & -40.7\%\\
makes independent decisions that affect our lives. & 11.1\% & 26.2\% & 11.1\% & 42.9\% & 37.5\% & -14.3\% & 33.3\% & -23.8\%\\
has its own consciousness. & -36.4\% & -22.2\% & 13.3\% & 11.1\% & 15.2\% & 44.4\% & -6.1\% & 33.3\%\\
determines our leisure time activities. & 23.3\% & 37.0\% & -20.0\% & -10.0\% & 30.0\% & -7.4\% & 26.7\% & -16.7\%\\
can optimize itself. & 9.5\% & 22.2\% & -14.3\% & 11.1\% & 61.9\% & 33.3\% & 28.6\% & 11.1\%\\
leads to personal loneliness. & -4.2\% & 8.3\% & -12.5\% & 16.7\% & 12.5\% & -33.3\% & 16.7\% & -33.3\%\\
controls what and how we learn. & 8.3\% & 20.8\% & 2.8\% & -4.2\% & 15.2\% & 25.0\% & 11.1\% & 8.3\%\\
becomes part of the human body. & 25.5\% & 37.5\% & 13.7\% & 33.3\% & 43.8\% & 33.3\% & 50.0\% & -33.3\%\\
has a sense of responsibility. & -19.0\% & -9.5\% & -23.1\% & 19.0\% & 15.2\% & 42.9\% & 17.9\% & 9.5\%\\
prevents crimes. & 25.9\% & 33.3\% & -11.1\% & 50.0\% & 40.7\% & 41.7\% & 25.9\% & 4.2\%\\
controls our search for partners. & 29.2\% & 33.3\% & 16.7\% & -44.4\% & 16.7\% & 33.3\% & 8.3\% & 0.0\%\\
is always subordinate to us in working life. & 5.6\% & 5.6\% & -33.3\% & 5.6\% & 46.7\% & 16.7\% & 55.6\% & 22.2\%\\
acts according to moral concepts. & -6.7\% & -7.4\% & 13.3\% & 77.8\% & 6.7\% & 18.5\% & 6.7\% & 3.7\%\\
learns faster than humans. & 60.6\% & 57.1\% & -13.9\% & 4.8\% & 48.5\% & 81.0\% & 44.4\% & 28.6\%\\
raises our standard of living. & 70.8\% & 66.7\% & 0.0\% & -33.3\% & 57.1\% & 71.4\% & 71.4\% & 28.6\%\\
controls hybrids of humans and technology. & -2.0\% & -6.7\% & 19.6\% & 13.3\% & 15.7\% & 6.7\% & 11.8\% & -3.3\%\\
is influenced by an elite. & 52.4\% & 47.1\% & 35.9\% & 54.9\% & 15.4\% & -29.4\% & -7.7\% & -35.3\%\\
supervises our private life. & 63.3\% & 57.6\% & 40.0\% & 72.7\% & -37.0\% & -18.2\% & -25.0\% & -63.6\%\\
supervises our behavior in public. & 64.1\% & 55.6\% & 23.1\% & 27.8\% & 41.0\% & 22.2\% & 2.6\% & -27.8\%\\
is misused by criminals. & 55.6\% & 46.7\% & 100.0\% & 40.0\% & -25.0\% & -20.0\% & -50.0\% & -46.7\%\\
is ahead of humans in its abilities. & 29.2\% & 20.0\% & 7.4\% & -26.7\% & 25.9\% & 44.4\% & 12.5\% & 6.7\%\\
is a family member. & 0.0\% & -13.3\% & -3.7\% & 0.0\% & 37.0\% & -33.3\% & 20.0\% & -13.3\%\\
creates many jobs. & 5.6\% & -8.3\% & -22.2\% & -38.9\% & 20.0\% & 50.0\% & 33.3\% & 47.2\%\\
contributes to solving complex social problems. & -7.7\% & -22.2\% & -7.7\% & 27.8\% & 15.4\% & -22.2\% & 33.3\% & -16.7\%\\
supports me as a helper in my tasks. & 63.6\% & 47.6\% & -30.0\% & -33.3\% & 46.7\% & 42.9\% & 60.0\% & 47.6\%\\
considers humans as a threat. & -22.9\% & -40.7\% & 6.2\% & 40.7\% & -8.9\% & -51.9\% & 2.2\% & -70.4\%\\
runs companies. & 0.0\% & -18.5\% & 16.7\% & 40.7\% & -8.3\% & -11.1\% & 8.3\% & -55.6\%\\
administers justice in legal matters. & -9.5\% & -30.3\% & 0.0\% & 51.5\% & 0.0\% & -21.2\% & 28.6\% & -54.5\%\\
knows everything about me. & 20.0\% & -4.8\% & 16.7\% & 52.4\% & 16.7\% & -33.3\% & -6.7\% & -33.3\%\\
controls what messages we receive. & 38.5\% & 8.3\% & 35.7\% & 44.4\% & -14.3\% & -33.3\% & -17.9\% & -50.0\%\\
occupies leadership positions in working life. & 6.7\% & -25.0\% & 13.3\% & 50.0\% & 13.3\% & 0.0\% & 3.3\% & -66.7\%\\
reduces our need for interpersonal relationships. & 18.5\% & -14.3\% & 18.5\% & 38.9\% & 7.4\% & -52.4\% & -11.1\% & -61.9\%\\
makes society lazy. & 33.3\% & 0.0\% & 0.0\% & 33.3\% & 36.1\% & 0.0\% & 19.4\% & -33.3\%\\
conducts international diplomacy. & 0.0\% & -33.3\% & 16.7\% & 33.3\% & 0.0\% & 16.7\% & 11.1\% & -33.3\%\\
destroys humanity. & -19.0\% & -54.2\% & 14.3\% & -20.8\% & -28.6\% & -20.8\% & -4.8\% & -37.5\%\\
is omniscient. & 0.0\% & -36.1\% & -4.2\% & 16.7\% & 28.6\% & 12.1\% & 25.0\% & -2.8\%\\
destroys many jobs. & 52.8\% & 16.7\% & 25.0\% & 44.4\% & 0.0\% & -27.8\% & 5.6\% & -38.9\%\\
independently writes scientific articles. & 54.5\% & 11.9\% & 3.0\% & -2.4\% & 15.2\% & 45.2\% & 30.3\% & 19.0\%\\
decides about our death. & -14.8\% & -63.3\% & 23.3\% & 40.0\% & 25.9\% & -63.0\% & 6.7\% & -83.3\%\\
makes our society more sustainable. & 45.5\% & -6.7\% & -27.3\% & -40.0\% & 60.6\% & 46.7\% & 63.6\% & 36.7\%\\
helps us to have better relationships. & 37.5\% & -14.8\% & -12.5\% & -11.1\% & 37.5\% & 7.4\% & 20.8\% & -22.2\%\\
\textbf{Average} & \textbf{11.9\%} & \textbf{22.2\%} & \textbf{3.5\%} & \textbf{18.4\%} & \textbf{21.5\%} & \textbf{7.1\%} & \textbf{19.7\%} & \textbf{-11.9\%}\\*
\insertTableNotes
\end{longtable}
\end{ThreePartTable}
\end{landscape}